\documentclass[pra,aps,epsf,twocolumn,floatfix]{revtex4}
\newcommand{\rem}[1]{}
\usepackage{graphicx}
\begin{document}
\title{
Entanglement between a qubit and the environment
in the spin-boson model
}
\author{T. A. Costi}
\email{tac@tkm.physik.uni-karlsruhe.de}
\affiliation
{Universit\"{a}t Karlsruhe, Institut f\"{u}r Theorie der Kondensierten
Materie, 76128 Karlsruhe, Germany}

\author{Ross H. McKenzie}
\email{mckenzie@physics.uq.edu.au}
\affiliation{ Department of Physics, 
University of Queensland, Brisbane, 4072,    Australia}

\begin{abstract}
The quantitative description of the quantum entanglement
between a qubit and its environment is considered.
Specifically, for the ground state of the spin-boson
model, the entropy of entanglement of the spin
is calculated as a function of $\alpha$, the strength
of the ohmic coupling to the environment, and $\varepsilon$, 
the level asymmetry.
This is done by a numerical renormalization group
treatment of the related anisotropic Kondo model.
For $\varepsilon=0$, the entanglement increases monotonically with
$\alpha$, until it becomes maximal for $\alpha\rightarrow 1^{-}$. For 
fixed $\varepsilon>0$,
the entanglement is a maximum as  a function of $\alpha$
for a value, $\alpha = \alpha_M < 1$.
\end{abstract}

\date{\today}

\maketitle


Due to the promise of quantum computation there
is currently considerable interest in the relationship
between entanglement, decoherence, entropy, and measurement.
Motivated by quantum information theory several authors
have recently investigated entanglement
in quantum many-body systems \cite{osborne,nature,hines,vidal}.
It is often stated that decoherence or a measurement
causes a system to become entangled with its environment.
The purpose of this paper is to make these ideas quantitative
by a study of the simplest possible model,
 the spin-boson model \cite{leggett.87,weiss}.
This describes a qubit (two-level system)
interacting with an infinite collection of harmonic oscillators
that model the environment responsible for decoherence
and dissipation. Specifically, we show how
the entanglement between a superposition state of the
qubit and the environment changes
as the coupling between the qubit and environment increases.
One interesting result is that we find that the qubit becomes
maximally entangled with the environment when the coupling $\alpha$
approaches a particular finite value ($\alpha\rightarrow 1^{-}$).
Furthermore, at this value    
the model undergoes a quantum phase transition, which
is consistent with recent observations that often
entanglement is largest near quantum critical
points\cite{osborne,nature,hines,vidal}.

{\em The spin-boson model.}
The Hamiltonian is
\cite{leggett.87,weiss}
\begin{eqnarray}
H_{SB} & = &\frac{1}{2}\Delta {\sigma}_{x}
+\frac{1}{2}\varepsilon{\sigma}_{z}
        +\sum_{i} \omega_{i}(a_{i}^{\dagger}a_{i}+\frac{1}{2})\nonumber\\
        &+&\frac{1}{2}{\sigma}_{z}\sum_{i}
\lambda_{i}(a_{i}+a_{i}^{\dagger})\label{eq:SB},
\end{eqnarray}
where $\Delta$ is the bare tunneling amplitude between the two quantum
mechanical states $\uparrow$ and $\downarrow$, $\varepsilon$ is the level
asymmetry (or bias),
 $\omega_i$ are the frequencies of the oscillators and $\lambda_i$ 
the strength with which they couple to the two quantum mechanical states.
The effect of the oscillator bath is completely determined by
the spectral function $J(\omega)$, defined below\cite{leggett.87}.
We will only consider the     
ohmic case, where it is has a linear dependence on
frequency  
$$J(\omega)=\pi\sum_{i}\lambda_{i}^{2}
\delta(\omega-\omega_{i})=2\pi \alpha \omega$$
for $\omega \ll \omega_{c}$, and $\alpha $ is the 
dimensionless dissipation strength.
The cutoff frequency, $\omega_{c} \gg \Delta$.   
This model can describe the decoherence of 
Josephson junction qubits, such as those recently
realized experimentally\cite{expt}, due
to voltage fluctuations in the electronic circuit\cite{schon},
and $\alpha$ can be expressed in terms of resistances
and capacitances in the circuit and so this is
an experimentally tunable parameter.
Recent results show it is possible to construct
devices, with $\alpha \ll 1$, the regime required
for quantum computation.
However, when modeling measurements one has $\alpha \sim 1$.

The dynamical properties of the model have been
extensively studied. In particular, suppose the spin (qubit) 
is initially in a pure state which is a product state
of up spin and the environment state,  then    the
coherent Rabi oscillations that would be observed
in the absence of coupling to the environment are modified
as follows.
One finds distinct behaviour for
$ 0 < \alpha < 1/2$ (damped coherent oscillations),
$ 1/2 < \alpha < 1$ (exponential decay),
and $1 < \alpha$ (localization; i.e., the spin remains in
the up state)
\cite{leggett.87,weiss,saleur,costi.96}.

{\em Entropy of entanglement.}
We now consider a quantitative description of the entanglement of
the qubit with
the environment. A good entanglement measure for a pure
state is the
von Neumann entropy or 
entropy of entanglement\cite{bennett,vedral}
\begin{equation}
E(\rho) = -Tr (\rho \log_2 \rho)
\label{entropy}
\end{equation}
where $\rho$ is the reduced density matrix of the qubit.
This is a two by two matrix given by 
\begin{equation}
\rho = {1 \over 2} \left( 1 + \sum_{a=x,y,z}
 \langle\sigma_a\rangle \sigma_a \right),
\label{rho}
\end{equation}
where $ \langle\sigma_a\rangle$ denotes the expectation value in
the state of interest.
In this case Eq.\ (\ref{entropy}) reduces to
\begin{equation}
E(\rho) =
-p_+  \log_2 p_+
-p_-  \log_2 p_-,
\label{entropy2}
\end{equation}
where $p_\pm$ are the eigenvalues of the density matrix,
\begin{equation}
p_\pm = {1 \over 2} \left( 1 \pm |\langle\vec{\sigma}\rangle|
\right).
\label{eigen}
\end{equation}

For $\varepsilon=0$ the only nonzero value of $ \langle\sigma_a\rangle$ is
$\langle\sigma_x\rangle$. At $T=0$ it is given by
\begin{equation}
 \langle\sigma_x\rangle = 2 {\partial E_{0} \over \partial \Delta},
\label{hellmann}
\end{equation}
where $E_{0}$ is the ground-state energy 
of the Hamiltonian (\ref{eq:SB}), and
use has been made of the Feynman-Hellmann theorem.
That the other values are zero can be seen by symmetry as follows.
In general the Hamiltonian is invariant under the 
reflection in spin space, $ \sigma_y \to - \sigma_y$.
Hence, all eigenstates must have a definite parity under
this transformation.
Thus, $ \langle\sigma_y\rangle=-\langle\sigma_y\rangle$ for all states and so
$\langle\sigma_y\rangle=0$  at any temperature. For $\varepsilon =0$ 
the Hamiltonian is also  invariant under the joint transformation 
$\sigma_z \to -\sigma_z$ and $a_i \to -a_i$ and so $\langle\sigma_z\rangle=0$ 
at any temperature, provided there is no symmetry breaking.

The challenge is now to evaluate the ground-state expectation values
$\langle\sigma_x\rangle$ and $\langle\sigma_z\rangle$.
For $\alpha > 1/2$ and particularly for $\alpha \sim 1$ 
this is a highly nontrivial
problem because in this regime nonperturbative
effects become important\cite{leggett.87,weiss}. However, we show how 
these expectation values can be evaluated  using
the numerical renormalization group (NRG) applied to the equivalent 
anisotropic Kondo model.

{\em Anisotropic Kondo Model.}
The above model is equivalent to the anisotropic Kondo 
model (AKM), defined by \cite{anderson.69}
\begin{eqnarray}
H &=& \sum_{k,\sigma} \epsilon_{k}c_{k\sigma}^{\dagger}c_{k\sigma}
+\frac{J_{\perp}}{2}
        (c_{0\uparrow}^{\dagger}c_{0\downarrow}S^{-} +
         c_{0\downarrow}^{\dagger}c_{0\uparrow}S^{+})
\nonumber\\
&+& \frac{J_{\parallel}}{2}
         (c_{0\uparrow}^{\dagger}c_{0\uparrow} -
          c_{0\downarrow}^{\dagger}c_{0\downarrow})S^{z} 
+ g\mu_{B}hS_{z}.\label{eq:AKM}
\end{eqnarray}
The first term represents a free electron conduction band. We use a 
flat density of states $\rho_{0}=1/2D_{0}$ per spin, with 
$2D_{0}$ the bandwidth. 
The second and third terms represent the
transverse and  longitudinal parts of the exchange interaction between a  
$S=1/2$ impurity and the local conduction-electron spin density, and
the last term represents a Zeeman term for a magnetic field 
coupling only to the impurity spin.
The correspondence between $H$ and $H_{SB}$, established via 
bosonization\cite{guinea.85,costi.99}, implies $\varepsilon=g\mu_{B}h$, 
$\frac{\Delta}{\omega_{c}}= \rho_{0} J_{\perp}$ and 
$\alpha=(1+ 2 \delta/ \pi)^{2}$, where
$\tan{\delta}= -\pi \rho_{0} J_{\parallel}/4$. $\delta$ is the phase 
shift for scattering of electrons from a potential $J_{\parallel}/4$
\cite{leggett.87,costi.96,guinea.85}. 
We choose $\omega_{c}=2D_{0}$ so that $\Delta = J_{\perp}$. 
This equivalence has been used extensively to make
predictions about the dynamics\cite{costi.96} and thermodynamics
\cite{costi.99} of the ohmic spin-boson model. The relevant
low energy scale for the thermodynamics is the Kondo scale 
$T_K(J_\perp,J_{\parallel})$ which is identified with the renormalized
tunneling amplitude, $\Delta_r$, of the spin-boson model,
\begin{equation}
{\Delta_r \over \omega_c} = \left({\Delta \over \omega_c}
\right)^{1/{(1-\alpha)}}.
\end{equation}
%
%
We restrict ourselves in this paper to the longitudinal 
sector of the AKM, i.e., to $J_\perp < |J_{\parallel}|$, where the
simple parameter correspondence between the models given above 
remains valid to lowest order in $\Delta/\omega_c = \rho_{0} J_{\perp}$. 
For larger values of $\Delta/\omega_c$, $\alpha$ will acquire a 
renormalization due to finite $\Delta=J_\perp$, as indicated by the 
scaling analysis of the 
AKM in Refs.\ \cite{anderson.69,costi.99}. This renormalization, however, is 
important mainly for the transverse sector of the AKM, 
$J_\perp > |J_{\parallel}|$, which we do not consider in this paper.

We turn now to the evaluation of $\langle \sigma_x \rangle$. 
The equivalence between models ensures that the AKM has (to within an additive
constant) the same ground-state energy $E_{0}$ as that of the spin-boson model.
At $T=0$, we therefore find, in analogy to Eq.\ (\ref{hellmann}) applied to 
the AKM with $\Delta=J_{\perp}$, that 
\begin{equation}
\langle \sigma_x \rangle =
\langle\ c_{0,\uparrow}^{\dagger}c_{0,\downarrow}S^{-}+H.c. \rangle
\label{identity}
\end{equation}
i.e., $\langle \sigma_x \rangle$ can be obtained from  
a {\em local} static correlation function. 
Another way of seeing that this relation is valid, is to note that 
the unitary transformation in bosonization which transforms $H$ 
into $H_{SB}$ also transforms 
$(c_{0,\uparrow}^{\dagger}c_{0,\downarrow}S^{-}+H.c.)$ into
$\sigma_x$ of the spin-boson model (details of this mapping can be found in
Ref.\ \cite{guinea.85} and in greater detail in Appendix A of 
Ref.~\onlinecite{costi.99}). The same unitary
transformation on the AKM transforms $\langle S_z \rangle$ into 
$\sigma_z/2$ of the spin-boson model. The latter can therefore 
be calculated directly within the AKM as a thermodynamic average 
$\langle 2S_z \rangle$.

{\em Method.}
The above local correlation function can be calculated from Wilson's NRG 
method \cite{wilson.75+kww.80} which has been shown to
give very reliable results
for quantum impurity models such as the AKM\cite{dmrg}.
 The approach used here allows in addition 
the calculation of local dynamical quantities, such as the dynamical
susceptibility $\langle\langle\ c_{0,\uparrow}^{\dagger}c_{0,\downarrow}S^{-};
c_{0,\downarrow}^{\dagger}c_{0,\uparrow}S^{+}\rangle\rangle$ 
\cite{costi.94}. In outline (see Ref.\ \cite{wilson.75+kww.80} 
for the details), 
the procedure consists of introducing a logarithmic mesh of $k$ points 
$k_{n}=\Lambda^{-n}, \Lambda > 1$ for the conduction band 
and performing a unitary transformation of the $c_{k\sigma}$ such
that $f_{0\sigma}=\sum_{k}c_{k\sigma}$ is the first operator in a new 
basis, $f_{n\sigma},n=0,1,\dots$, which tridiagonalizes 
$H_{c}=\sum_{k\mu}\epsilon_{k\mu}c_{k\mu}^{\dagger}c_{k\mu}$ 
in k space. The Hamiltonian (\ref{eq:AKM}) with the new discretized 
form of the kinetic energy is now diagonalized by the following iterative 
process: (a) one defines a sequence of finite size Hamiltonians 
$H_{N} = \sum_{\mu}\sum_{n=0}^{N-1}\xi_{n}\Lambda^{-n/2}
(f_{n+1\mu}^{\dagger}f_{n\mu}+ H.c.) + 
\frac{J_{\perp}}{2}
        (f_{0\uparrow}^{\dagger}f_{0\downarrow}S^{-} +
         f_{0\downarrow}^{\dagger}f_{0\uparrow}S^{+})
  + \frac{J_{\parallel}}{2}
         (f_{0\uparrow}^{\dagger}f_{0\uparrow} -
          f_{0\downarrow}^{\dagger}f_{0\downarrow})S^{z}$ for $N\ge 0$ and 
$\xi_n \rightarrow 1$ for $n\gg 1$ \cite{wilson.75+kww.80}; 
(b) the sequence
of Hamiltonians ${H}_{N}$ for $N=0,1,\ldots$ is iteratively diagonalized
within a product basis of, typically, up to 1200 states for each iteration,
up to a maximum value $N=N_m$.
This gives the excitations and many body eigenstates at a corresponding
set of energy scales $\omega_{N}$ defined by the lowest scale 
$\omega_{N}=\Lambda^{-\frac{N-1}{2}}$ in $H_N$. The matrix elements 
$\langle m|O_{x,z}|n\rangle_{N}$ for the operators 
$O_x=c_{0,\uparrow}^{\dagger}c_{0,\downarrow}S^{-}$ and $O_z=S_{z}$, 
required to calculate $\langle\sigma_x\rangle$ and $\langle\sigma_z\rangle$, 
are also calculated iteratively. The choice of $N_m$ depends on the Kondo
scale $T_{K}=\Delta_r$ and hence on $\alpha$, but for given $\alpha$ (i.e.
for given $J_{\perp},J_{\parallel}$) should be large 
enough such that $\omega_{N_m}\ll \Delta_r$. A discretization parameter
$\Lambda=1.5$ was used throughout and we checked that the above expectation
values remained unchanged on further increasing $N_m$. This suggested that
our approximation of using a finite $N_m$ to calculate the thermodynamic
expectation values of the infinite system is a very good one.
\begin{figure}[t]
\includegraphics[width=8cm]{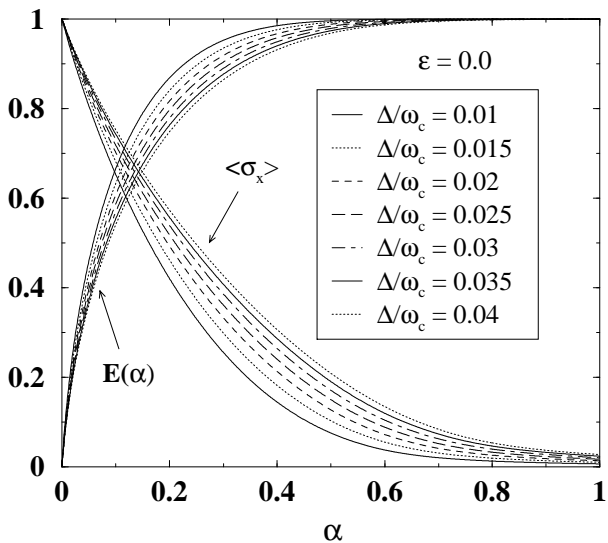}
\vspace{0.1cm}
\caption{
The dependence of (i) the ground-state expectation value
$\langle\sigma_x\rangle$ 
as a function of the dimensionless coupling $\alpha$ to the environment for
$\varepsilon = 0$, and (ii) the entanglement entropy $E$ 
of a qubit ohmically coupled to an environment as
a function of $\alpha$ for $\varepsilon = 0$.
The different curves correspond to different
values of the ratio of the bare 
tunneling amplitude $\Delta$ to 
the cutoff frequency of the boson bath $\omega_{c}$.
Note that as $\alpha \to 1^-$
the qubit becomes maximally entangled with the environment.
}
\label{sigmax+entropy-eps0}
\end{figure}

{\em Symmetric case.}
For $\varepsilon=0$ and $\alpha < 1$ only $\langle\sigma_x\rangle$ is 
nonvanishing. We show
this in Fig.\ (\ref{sigmax+entropy-eps0}) at $T=0$
versus dimensionless dissipation strength $\alpha$ in the range 
$0 < \alpha < 1$, and for several values of 
the dimensionless tunneling amplitude $\Delta/\omega_c$.
The limiting noninteracting value $\langle\sigma_x\rangle\rightarrow 1$ 
is recovered as $\alpha\rightarrow 0$. In the limit 
$\Delta/\omega_c \rightarrow 0$ it vanishes at the quantum critical 
point of the spin-boson model $\alpha = 1$ where $\Delta_r\rightarrow 0$. 
For any finite fixed $\Delta/\omega_c$, however,  our use of the AKM implies
that the critical behaviour occurs at $\alpha_{c}>1$ with 
$\alpha_{c}\rightarrow 1$ as $J_{\perp}\rightarrow 0$ (specifically, this
critical behaviour occurs at the ferromagnetic-antiferromagnetic 
boundary $J_{\perp}=-J_{\parallel}$).
Figure (\ref{sigmax+entropy-eps0}) also shows the $T=0$ entropy 
of entanglement of the qubit. The entropy vanishes as 
$\alpha\rightarrow 0$ and approaches
its maximum value as $\alpha\rightarrow 1^{-}$ 
(see also Ref.\ \cite{sidles.03} for weak
dissipation results for E). For $\alpha > 1$,
we are in the ferromagnetic sector of the AKM where $\langle\sigma_z\rangle=1,
\langle\sigma_x\rangle=0$, and the reduced density matrix 
eigenvalues $p_{\pm}=0,1$ giving $E=0$, i.e.,
$E(\alpha)$ drops discontinuously at the quantum critical point $\alpha=1$ 
\cite{long-paper}. It is interesting that for a spin qubit 
coupled to two bosonic baths it is possible to remain in the delocalized
phase (i.e., $\langle\sigma_z\rangle = 0$) for all dissipation strengths 
\cite{castro-neto.03}.
Finally, we note that the entropy of entanglement is quite different from
the thermodynamic entropy of the boundary
(or impurity spin entropy). The latter  is usually defined as
$S(\alpha) - S(\alpha=0)$ where $S(\alpha)$ is the total
thermodynamic entropy of the system \cite{costi.99}.
The impurity spin entropy is zero for $\alpha < 1$
because the ground state of the AKM is a spin singlet for 
$J_\parallel>0$. 
\begin{figure}[t]
\includegraphics[width=8cm]{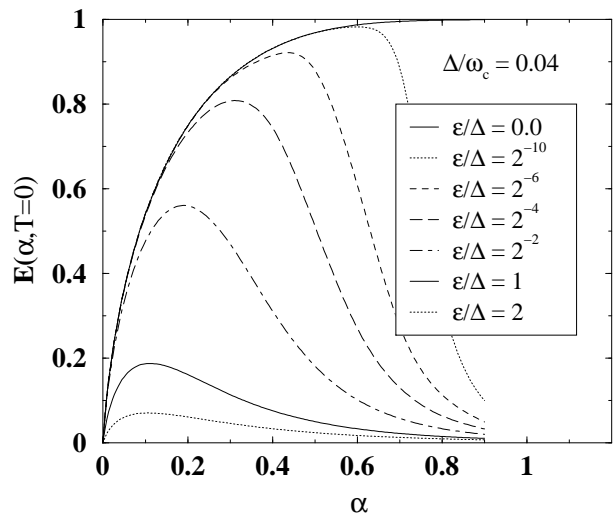}
\vspace{0.1cm}
\caption{
The dependence of the  entanglement entropy 
of the ground state
on the coupling to the environment $\alpha$ and 
the level asymmetry $\varepsilon$ for $\Delta/\omega_{c}=0.04$.
Note that for $\varepsilon>0$, the entanglement is a maximum
at $\alpha=\alpha_M < 1$.}
\label{entropy-t0-asym}
\end{figure}

\begin{figure}[t]
\includegraphics[width=8cm]{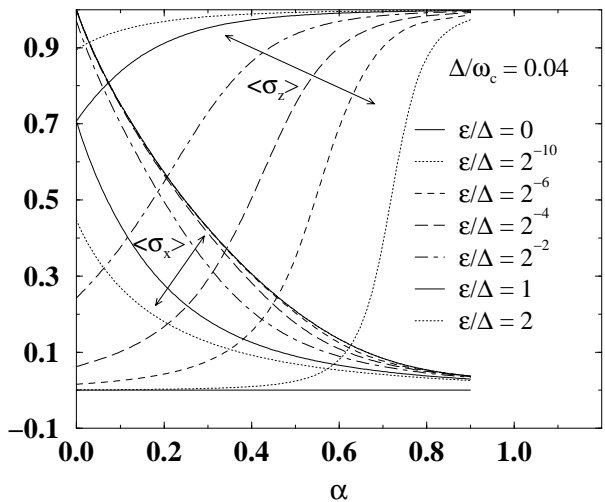}
\vspace{0.1cm}
\caption{
The dependence of the ground state expectation values
$\langle\sigma_x\rangle$, $\langle\sigma_z\rangle$
on $\alpha$ and $\varepsilon$ for $\Delta/\omega_{c}=0.04$.
For $\alpha=0$ the noninteracting values
 $\langle\sigma_x\rangle=\Delta/\sqrt{\varepsilon^2 + \Delta^2}$
and $\langle\sigma_z\rangle=\varepsilon/\sqrt{\varepsilon^2 + \Delta^2}$
 are recovered for all values of $\varepsilon/\Delta$.
}
\label{sigmaxz}
\end{figure}

{\em Asymmetric case.}
For $\varepsilon > 0$, $\langle\sigma_z\rangle$ acquires a
finite value analogous to the 
magnetization $\langle S_z\rangle$ in a local 
magnetic field $g\mu_{B}h=\varepsilon$ in the AKM.
The entanglement entropy $E$ now depends on $|\langle\vec{\sigma}\rangle| = 
(|\langle\sigma_x\rangle|^2+|\langle\sigma_z\rangle|)^{1/2}$ via
Eq.\ (\ref{rho}) and is shown in Fig.\ (\ref{entropy-t0-asym}). 
The behaviour of $E$ as a function of $\alpha$ and $\varepsilon$ 
is understood from the behaviour of  $\langle\sigma_x\rangle$ and  
$\langle\sigma_z\rangle$ shown in  Fig.\ (\ref{sigmaxz}).
In particular, we now find that for arbitrary small $\varepsilon$, 
the entanglement entropy first increases with increasing $\alpha$ 
before reaching a maximum value
at $\alpha=\alpha_M < 1$ and then decreasing as $\alpha\rightarrow 1$.
This behaviour arises from the competition between the 
$\alpha$ dependence of $\langle\sigma_x\rangle$ and 
$\langle\sigma_z\rangle$ in Fig.\ (\ref{sigmaxz}). Whereas 
$\langle\sigma_x\rangle$ 
continues to decrease monotonically with increasing $\alpha$ 
(as for $\varepsilon=0$), it is seen that 
$\langle\sigma_z\rangle$ increases 
monotonically with increasing $\alpha$ with 
$\langle\sigma_z\rangle\rightarrow 1$ as $\alpha\rightarrow 1$. 
The condition for full polarization, $\langle\sigma_z\rangle\approx 1$, is
$\varepsilon \gg \Delta_r$. For any $\varepsilon>0$, this condition 
is always satisfied since $\Delta_r\rightarrow 0$ as $\alpha\rightarrow 1$. 
It follows that $|\langle\vec{\sigma}\rangle|$ has a minimum 
as a function of $\alpha$ and that the entanglement 
entropy, for $\varepsilon>0$, has a maximum at $\alpha=\alpha_M <1$ 
before decreasing again as $\alpha\rightarrow 1$.

Finally, we suggest several directions
for future work.

(i) This work focused solely on static properties of 
the spin-boson model.
It would be interesting to consider dynamics, for example, the 
longitudinal and the transverse dynamical susceptibilities, and 
hence extract the decoherence and relaxation
rates for an ohmically coupled qubit.
In addition, it is interesting to ask how the entanglement 
varies with time if the initial state has no entanglement 
of the qubit and the environment. 

(ii) 
The AKM is integrable by the Bethe ansatz\cite{tsvelick.83}.
The AKM can  also be related to a free boson field theory
with a boundary sine-Gordon term\cite{saleur,saleur2}
which is also integrable by the Bethe ansatz.
Exact expressions can be obtained for the free energy. It involves
solving a set of thermodynamic Bethe ansatz
(TBA) equations. At $T=0$ the impurity ground state
energy is going to be related to $T_K$. The real problem
is getting results for arbitrary anisotropies 
(dissipation strengths)\cite{TBA}.

(iii) Recently it was shown\cite{vidal} that if in a 
quantum critical system one calculates
the entropy of entanglement of
a subsystem of size $L$ with the rest of
the system this equals the geometric
entropy previously calculated
for the corresponding conformal field theory
(motivated by questions concerning black
hole thermodynamics!)\cite{holzhey}.
It would be interesting to perform similar calculations
for the relevant boundary field theory.

(iv) The NRG can also be used to reliably calculate
properties of the spin-boson model at nonzero temperature\cite{long-paper}.
However, calculating the entanglement at nonzero temperature
is an open problem because it involves a mixed state
and it is not practical to evaluate the measure
of entanglement that has been proposed for such
states\cite{bennett}.

T.A.C. acknowledges support from the Deutsche Forschungsgemeinschaft through
the Sonderforschungsbereich 195.
Work at UQ was supported by the Australian Research Council.
We  thank A. Hines, A. W. W. Ludwig, G. J. Milburn, and H. -Q. Zhou for 
very helpful discussions.



\begin{thebibliography}{99}
\bibitem{TBA}
The TBA equations are quite complicated for 
arbitrary anisotropies. For special anisotropies corresponding 
to dissipation strengths $\alpha=1/\nu$ and $\alpha=1-1/\nu$, $\nu=3,\dots$, 
the TBA equations are manageable and have
been solved previously\cite{costi.99}.
Going beyond this is very complicated, but technically 
possible \cite{costi.99}.
\end{thebibliography}
\end{document}